\documentclass[runningheads]{llncs}
\usepackage{graphicx}
\usepackage{cite}
\usepackage{amsmath,amssymb,amsfonts}
\usepackage{algorithm}
\usepackage{algpseudocode}
\usepackage{multirow}

\usepackage{subcaption}

\usepackage{xcolor}
\usepackage{acronym}

\newacro{sca}[SCA]{Side-Channel Attack}
\newacro{scca}[SCCA]{Side-Channel Cryptographic Attack}
\newacro{snr}[SNR]{Signal to Noise Ratio}
\newacro{poi}[POI]{Point of Interest}
\newacroplural{poi}[POIs]{Points of Interest}
\newacro{ml}[ML]{Machine Learning}
\newacro{dl}[DL]{Deep Learning}
\newacro{rf}[RF]{Radio Frequency}
\newacro{aes}[AES]{Advanced Encryption Standard}
\newacro{sdr}[SDR]{Software Defined Radio}
\newacro{co}[CO]{Cryptographic Operation}
\newacro{cp}[CP]{Cryptographic Process}
\newacroplural{cp}[CPs]{Cryptographic Processes}
\newacro{pge}[PGE]{Partial Guessing Entropy}
\newacro{dpa}[DPA]{Differential Power Analysis}
\newacro{cpa}[CPA]{Correlation Power Analysis}
\newacro{mia}[MIA]{Mutual Information Analysis}
\newacro{em}[EM]{Electromagnetic}
\newacro{pw}[PW]{power supply}
\newacro{fc}[FC]{Frequency Component}
\newacro{da}[DA]{Differential Analysis}
\newacro{ca}[CA]{Correlation Analysis}
\newacro{ta}[TA]{Template Attacks}
\newacro{puf}[PUF]{Physical Unclonable Function}
\newacro{fpga}[FPGA]{Field-Programmable Gate Array}
\newacro{soc}[SoC]{System-on-a-Chip}
\newacro{iv}[IV]{Intermediate Value}
\newacro{bc}[BC]{Bruteforce Complexity}
\newacro{vt}[VT]{Virtual Trigger}
\newacro{tvla}[TVLA]{Test Vector Leakage Assesment}

\usepackage{scalerel}

\begin{document}
\title{Attacking at non-harmonic frequencies in screaming-channel attacks}

\author{Jeremy Guillaume\inst{1}\orcidID{0009-0005-3398-3423} \and 
Maxime Pelcat\inst{2}\orcidID{0000-0002-1158-0915} \and
Amor Nafkha\inst{1}\orcidID{0000-0002-1164-7163} \and
Rubén Salvador\inst{3}\orcidID{0000-0002-0021-5808}
}

\authorrunning{J. Guillaume et al.}
\institute{CentraleSupélec, IETR UMR CNRS 6164, France \and
Univ Rennes, INSA Rennes, CNRS, IETR - UMR 6164, F-35000 Rennes, France \and
CentraleSupélec, Inria, Univ Rennes, CNRS, IRISA, France}
\maketitle              
\begin{abstract}
Screaming-channel attacks enable \ac{em} \acp{sca} at larger distances due to higher \ac{em} leakage energies than traditional SCAs, relaxing the requirement of close access to the victim. 
This attack can be mounted on devices integrating \ac{rf} modules on the same die as digital circuits, where the \ac{rf} can unintentionally capture, modulate, amplify, and transmit the leakage along with legitimate signals.
Leakage results from digital switching activity, so previous works hypothesized that this leakage would appear at multiples of the digital clock frequency, i.e., harmonics.

This work demonstrates that compromising signals appear not only at the harmonics and that leakage at non-harmonics can be exploited for successful attacks. 
Indeed, the transformations undergone by the leaked signal are complex due to propagation effects through the substrate and power and ground planes, so the leakage also appears at other frequencies.
We first propose two methodologies to locate frequencies that contain leakage and demonstrate that it appears at non-harmonic frequencies.
Then, our experimental results show that screaming-channel attacks at non-harmonic frequencies can be as successful as at harmonics when retrieving a 16-byte AES key. 
As the \ac{rf} spectrum is polluted by interfering signals, we run experiments and show successful attacks in a more realistic, noisy environment where harmonic frequencies are contaminated by multi-path fading and interference.
These attacks at non-harmonic frequencies increase the attack surface by providing attackers with more potential frequencies where attacks can succeed.

\keywords{Cybersecurity \and hardware security \and electromagnetic side channels \and screaming-channel attacks}
\end{abstract}

\acresetall

\section{Introduction}
\acfp{sca}~\cite{choiTEMPESTComebackRealistic2020, standaert_IntroductionSideChannelAttacks_2010} allow retrieving confidential information from computing devices by exploiting the correlation of internal data with the leakage produced while computing over these data.
The term \emph{side channel} is therefore used to refer to physical leakage signals carrying confidential information. Side channels are general to CMOS computing devices and can take many forms, from runtime variations of system power consumption~\cite{mangard2008power} to \acf{em} emanations\cite{gandolfi_EM_2001}. 
Screaming channels are a specific form of \ac{em} side channel that occurs on mixed-signal devices, where a \ac{rf} module is co-located on the same die as digital modules. 
In this context, the leakage of the digital part reaches the \ac{rf} module, which can transmit it over a distance of several meters. This phenomenon allows attackers to mount side-channel attacks at distances from the victim. 
The seminal work of Camurati et al.~\cite{camurati_ScreamingChannelsWhen_2018} demonstrated how screaming-channel attacks can succeed at distances of up to 15 meters. 

Leakage is generated by the switching activity of the transistors from the digital part of the victim system, which operates at a clock frequency $F_{clk}$. When observed on a spectrum analyzer, the leakage power spectral density is shaped as peaks at the harmonics of $F_{clk}$ ($i.e.$ $n\times F_{clk}$ where $n \in  \mathbb{Z}$). 
What makes screaming-channel attacks different from other \acp{sca} is that the harmonics, after being modulated by the \ac{rf} module, are visible around the carrier frequency $F_{RF}$ of the legitimate \ac{rf} signal (Section~\ref{sec:screaming_channel_attacks}).

A limitation of this attack is that the harmonics of $F_{clk}$ can be modulated at the same frequency as some interfering signals, such as WiFi signals. Since these interfering signals are transmitted voluntarily, they are stronger than the leakage signal, which, as a result, can be easily polluted and hence quickly become non-exploitable.

To overcome this limitation and further study the risk posed by screaming channels, this paper studies the attack's feasibility when capturing signals at frequencies other than the harmonics of the digital processing clock.  
Specifically, we seek to answer the following questions: \textbf{is exploitable leakage also present at frequencies other than the harmonics? In case it is, is the difficulty for a successful attack higher at non-harmonics than at harmonics frequencies?} 
If the first question is answered positively, attackers can have an extensive choice of potential frequencies to select from and find one not polluted by environmental noise during the attack. 
Such a property can also be an enabler to effectively extend the framework of multi-channel attacks~\cite{Agrawal_multichannelAttack_2003}, which attack by combining different side-channel sources and different frequencies in the context of modulated leakage signals.

To summarize, we investigate the presence of leakage over the spectrum at non-harmonic frequencies and demonstrate that this leakage can be used to build successful attacks. We propose the following contributions:
\begin{itemize}
\item \textbf{Two methodologies to search for exploitable leakage over the spectrum}. The first is based on a \emph{fixed vs. fixed} t-test~\cite{durvaux2016improved}, and the second is an original contribution based on \textbf{\acf{vt}}~\cite{guillaume_VirtualTriggeringTechnique_2022}.
\item With these methodologies, we \textbf{demonstrate that leakage in screaming-channel attacks is not only present at harmonic frequencies}, as explored in previous works~\cite{camurati_ScreamingChannelsWhen_2018, wang_FarFieldEM_2020}, but it is also spread over a large share of the near-carrier spectrum.
\item We compare both  methods and demonstrate a \textbf{significant reduction in the exploration time} when looking for exploitable frequencies with the second methodology \textbf{based on pattern detection}. 
\item We evaluate the \textbf{effectiveness of attacks at non-harmonic frequencies} in a noiseless environment and show how this effectiveness can sometimes be higher at non-harmonics. 
\item We demonstrate \textbf{successful attacks in more realistic scenarios}. We apply the proposed methodologies, and the insights learned when attacking at non-harmonics and build attacks in a context where most harmonics are polluted by other standard signals typically found in the spectrum.
\end{itemize}

The rest of this paper is organized as follows: 
section~\ref{sec:RelatedWorks} introduces related and previous works on screaming-channel attacks. 
The attack scenario of our work and the setup are described in section~\ref{sec:AttackScenario}. 
The two methods we propose to search for leakage over the spectrum are presented in section~\ref{sec:SearchingLeakage}. 
Afterward, in section~\ref{sec:Attack_non_harmonics_wired}, we demonstrate the attack feasibility by exploiting the leakages found at non-harmonic frequencies.
Section~\ref{sec:Attack_challenging} demonstrates the attack in a more challenging, and therefore more realistic, scenario. 
Lastly, section~\ref{sec:Conclusion} concludes the paper.

\section{Related works}
\label{sec:RelatedWorks}
\subsection{Side-channel attacks}
By capturing the leakage generated by computing devices, attackers can mount \acp{sca} to jeopardize confidentiality and recover internal secret data.
The most common methods used to build \acp{sca} are 
\ac{dpa}\cite{kocher_DifferentialPowerAnalysis_}, 
\ac{cpa}\cite{brier_CorrelationPowerAnalysis_2004},  
\ac{mia}\cite{gierlichsMutualInformationAnalysis2008a}, 
\ac{ta}\cite{chari_TemplateAttacks_2003} 
and more recently \ac{dl}~\cite{masure_ComprehensiveStudyDeep_2019}.
The leakage signal of interest is generated by the digital part of the system from the switching activity of the transistors occurring when data is being computed or moved through the chip. 
The steep peaks of the signals resulting from the switching activity of transistors in digital devices produce leakage signals whose variations correlate with that switching activity and, therefore, with the data over which the processor computes.
That leakage signal is most often found in the \ac{em} emanations~\cite{gandolfi_EM_2001} or the power consumption variations~\cite{mangard2008power} of the victim device.

\begin{figure}[tb]
\centerline{\includegraphics[width=0.6\textwidth]{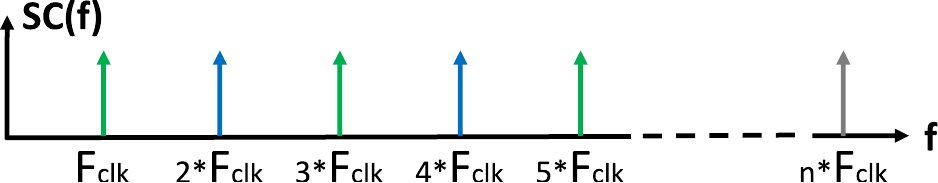}}
\caption{\textbf{Leakage presence over the spectrum:} The Fourier transform of a signal having a period Tclk consists of peaks at both odd harmonics (green peaks) and even harmonics (blue peaks) of frequency $F_{clk}$ = $1/T_{clk}$.}
\label{Harmonics}
\end{figure}

In a synchronous device, transistors switch at the pace of the digital clock. As a result, the leakage resulting from successions of transitions and non-transitions has a period equal to one clock cycle.
The Fourier Transform of a signal having a period $T$ corresponds to peaks at each harmonic of the frequency $F$ = $1/T$~\cite{adamczyk2017foundations} as formalized in Equation~\eqref{eq_H}, with $A_{n}$ being the respective amplitudes of each harmonic.
Then, theoretically, the  leakage should appear at harmonics of $F_{clk}$ = $1/T_{clk}$ as formalized in Equation~\eqref{eq_SC} and illustrated in Fig.~\ref{Harmonics}.
This corresponds to what is empirically observed in works on side-channel attacks where the leakage is found at harmonics of the clock frequency~\cite{agrawal_EMSideChannel_2003}.
\begin{equation}
\begin{split}
& H(f) = \Sigma_{n=-\infty}^{\infty} A_{n}\delta(f - n / T)\label{eq_H}
\end{split}
\end{equation}

\begin{equation}
 SC(f) =
     \Sigma_{n=-\infty}^{\infty} A_{n}\delta(f - nF_{clk})\label{eq_SC}
\end{equation}

A limitation of traditional \acp{sca} to capture clean leakage signals is that attackers must be in very close proximity to the victim device for a successful attack, usually only a few millimeters away. However, some specific scenarios allow attackers to take distance from the device~\cite{schellenberg_JobRemotePower_2018a, dessouky_SoKSecureFPGA_2021}. One scenario for gaining distance from the device, which is the focus of this paper, is the so-called screaming-channel attack~\cite{camurati_ScreamingChannelsWhen_2018}. In this attack, the leakage is transmitted by an \ac{rf} module that sits beside the digital part of a mixed-signal chip on the same die, allowing the attacker to capture it at a distance of several meters. 

\subsection{Screaming-channel attacks}
\label{sec:screaming_channel_attacks}
Mixed-signal devices are heterogeneous platforms with digital and analog modules integrated into the same die. One of these analog modules can be the \ac{rf} chain needed to build a \ac{soc} with radio communications capabilities. This tight integration has the advantage of reducing the power consumption or the transmission delay between the digital and \ac{rf} modules, as well as the cost of the final device. 
However, the very nature of this type of device has already been proven to be a hardware vulnerability~\cite{camurati_ScreamingChannelsWhen_2018}.

\begin{figure}[tb]
\centerline{\includegraphics[width=0.7\textwidth]{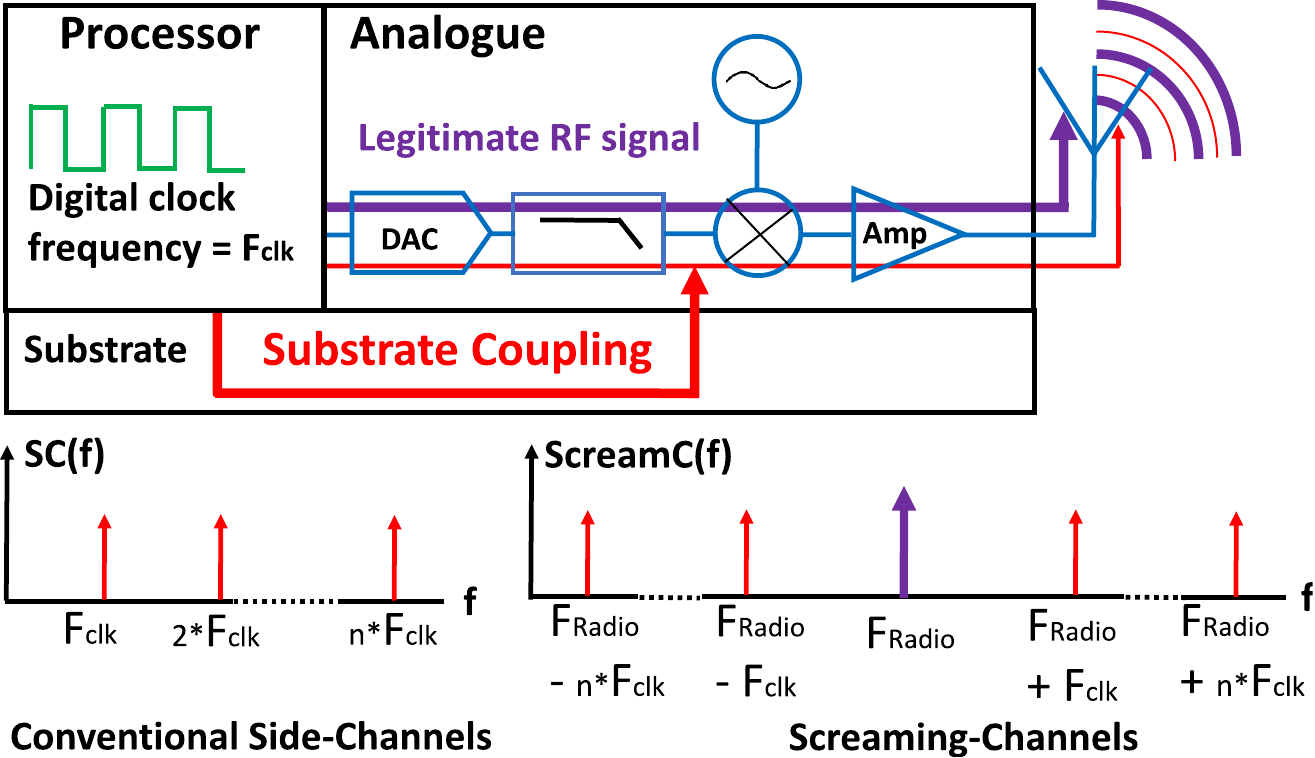}}
\caption{\textbf{Screaming-channel attacks:} The conventional leakage of the digital part perturbates the \ac{rf} module of the analog part, present on the same die. The radio transceivers here modulate the leakage around the frequency of the legitimate signal and transmit it at a longer distance (up to several meters).} 
\label{Screaming_Channels}
\end{figure}

Fig.~\ref{Screaming_Channels} illustrates a mixed-signal device. Compared to regular side-channels in digital devices, the leakage resulting from the switching activity in mixed-signal systems can travel 
through the substrate by the so-called \emph{substrate coupling} effect~\cite{le2011experimental, mohamed2010physical, rhee2008experimental, afzali2006substrate}. This way, leakage signals can reach the radio transceivers of the \ac{rf} part, which is very sensitive to noise and hence prone to capture these slight variations carrying information that correlates with secret data.

The leakage from the digital part, which is the one that would be used to build a traditional \ac{sca} is, as expressed in Equation~\eqref{eq_screamC}, modulated at the frequency of the legitimate \ac{rf} signal $F_{RF}$, amplified and then transmitted by the \ac{rf} module through the antenna. As a result, this amplification can bring, i.e., \emph{scream}, the leakage signal at distances of several meters.
In Camurati et al. use case~\cite{camurati_ScreamingChannelsWhen_2018}, the transmitted Bluetooth signal is centered at $2.4$GHz, and the device clock frequency is $64$MHz. 
The leakage harmonics are therefore at $2.4$GHz + multiples of $64$GHz, i.e., $2.464$GHz, $2.528$GHz, $2.592$GHz, etc.
\begin{equation}
 ScreamC(f) =  
     SC(f) \ast F_{RF} =
     \Sigma_{n=-\infty}^{\infty} A_{n}\delta(f - nF_{clk} - F_{RF})\label{eq_screamC}
\end{equation}
In their seminal work, Camurati et al.~\cite{camurati_ScreamingChannelsWhen_2018} demonstrated that an attack using this modulated leakage is possible. Authors also conducted further works to understand better the properties of the leakage in this context~\cite{camurati_UnderstandingScreamingChannels_2020}. In this original use case, the device transmits a Bluetooth signal from a commodity low-power \ac{soc} while the digital part (an Arm Cortex-M4 microcontroller) executes AES encryptions. The authors reported performing the screaming-channel attack at up to 15 meters and observed some leakage at up to 60 meters.

The limitation of this scenario is that the leakage carries low energy compared to other legitimate signals potentially present at the harmonic frequencies.
As a result, attacks can be very difficult to conduct at those frequencies when they are polluted.  
This sets a high uncertainty on the feasibility of the attack
in a noisy environment where all harmonics would be polluted.
The following sections demonstrate that attackers do not have to limit themselves to attack at a very limited set of harmonic frequencies. On the contrary, a wide spectrum is at their disposal to compromise the system, which increases the threat that screaming-channel attacks represent. To the best of our knowledge, this is the first work to demonstrate that the attack is possible at non-harmonics. All previous works on screaming-channel attacks~\cite{camurati_ScreamingChannelsWhen_2018, camurati_UnderstandingScreamingChannels_2020, wang_FarFieldEM_2020} used one harmonic (the second harmonic at $2.528$~GHz) to perform the attack.

\section{The attack scenario and setup}
\label{sec:AttackScenario}
This paper considers a scenario where the victim is a mixed-signal device that executes a \acf{cp} while transmitting an \ac{rf} signal in parallel. Since the leakage is emitted by the \ac{rf} module, the attacker can capture it with a \ac{sdr} device. The attacker's goal is to recover the secret key used by the \acp{cp} from this remotely captured leakage signal.

\begin{figure}[!tb]
    \centering
    \begin{subfigure}[b]{0.31\textwidth}
        \centering
        \includegraphics[width=\textwidth]{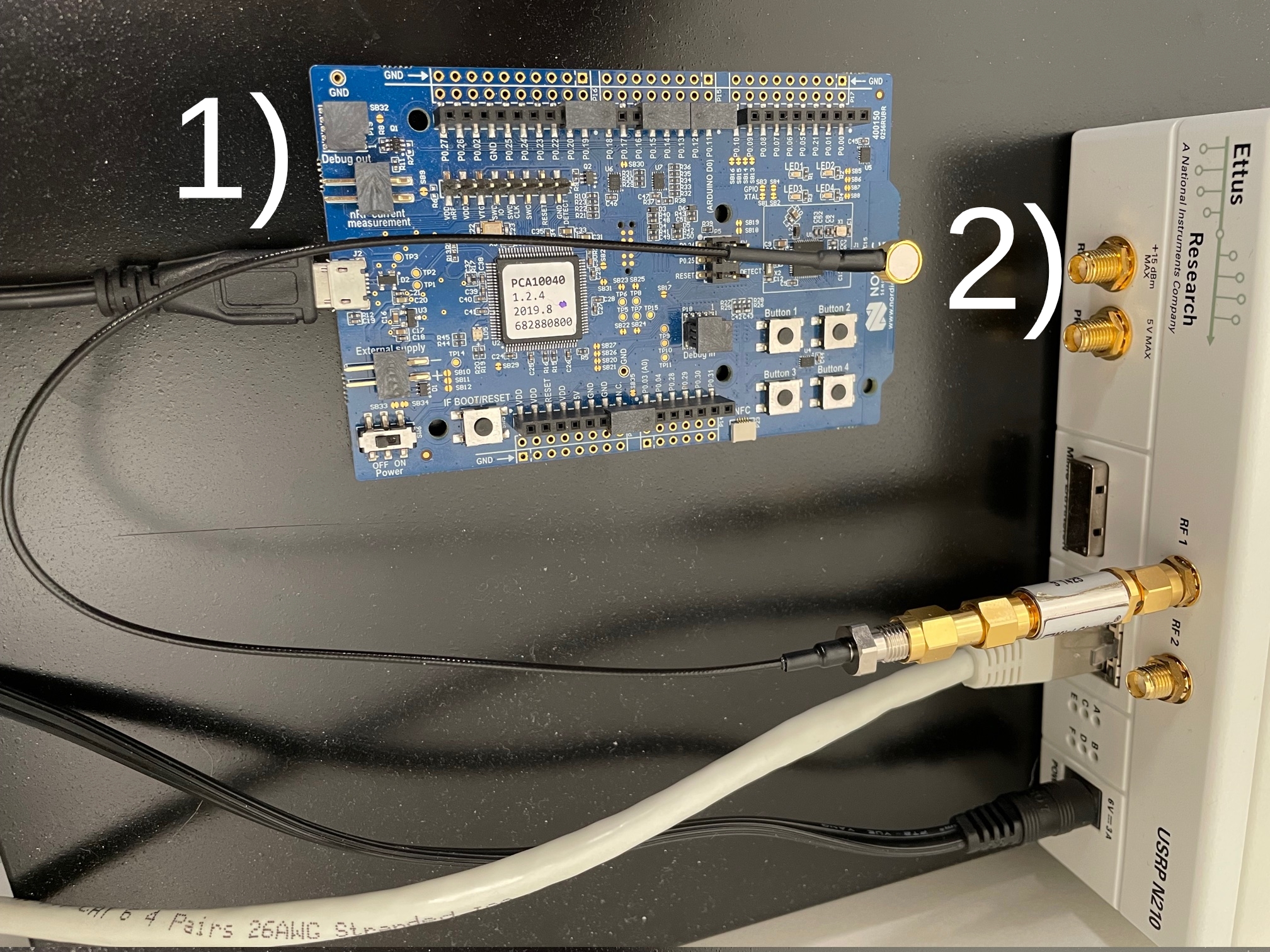}
        \caption{Wired setup}
        \label{fig:wired_setup}
    \end{subfigure}
    \hspace{6ex}
    \begin{subfigure}[b]{0.5\textwidth}
        \centering
        \includegraphics[width=\textwidth]{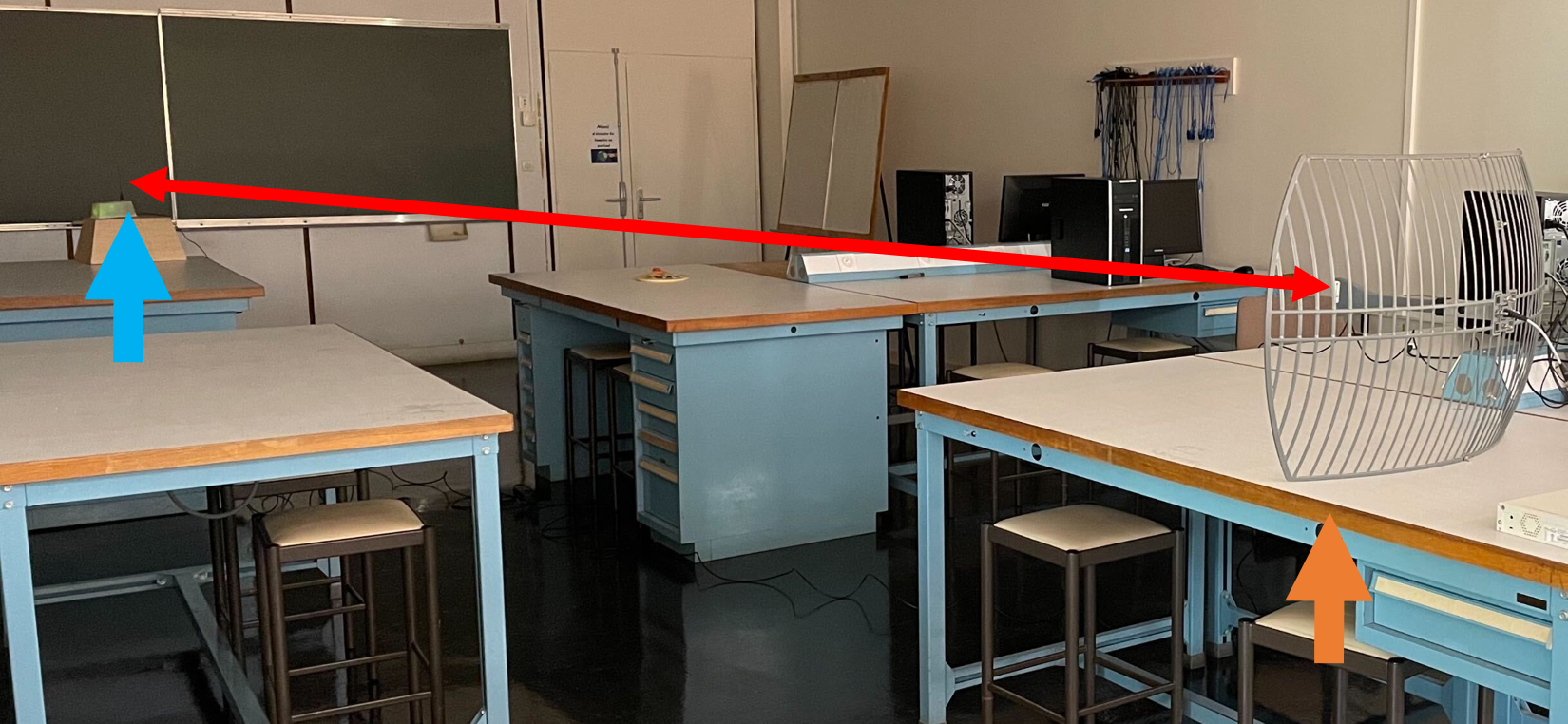}
        \caption{Wireless setup}
        \label{fig:setup_7meters}
    \end{subfigure}
    \caption{\textbf{Experimental setup:} (\subref{fig:wired_setup}) shows our wired setup for noiseless experiments and (\subref{fig:setup_7meters}) an attacker at 7 meters from the victim. This last one contains: 1) (blue arrow) the victim PCA10040 device running AES and transmitting a Bluetooth signal at $2.4$~GHz.; and 2) (orange arrow) the \ac{sdr} device that collects the data-correlated leakage from the victim.} 
    \label{Setup}
\end{figure}

Fig.~\ref{Setup} shows the attack setup used for all experiments.
The victim device is an nRF52832 from Nordic instrument\footnote{\url{https://www.nordicsemi.com/Software-and-tools/Development-Kits/nRF52-DK.}}. It contains an Arm Cortex M4 processor and an \ac{rf} module. 
The attacker device is a USRP N210\footnote{\url{https://www.ettus.com/all-products/un210-kit/}} using an SBX daughter board that can measure a signal between $400$~MHz and $4.4$~GHz and has a bandwidth of $40$~MHz. 
To collect the leakage at a distance, a parabolic grid  antenna with a gain of $26$~dBi is used.
The computer used for all our experiments has a 4-core Intel Xeon(R) CPU E3-1226 V3 @ 3.30 GHz, and $8$~GB RAM.

The legitimate \ac{rf} signal is a Bluetooth signal transmitted at $2.4$GHz without frequency hopping\footnote{\url{Frequency-hopping is the repeated switching of the carrier frequency during radio transmission to reduce interference and avoid interception. In the case of Bluetooth transmissions, switching occurs among $81$ channels, from $2.4$GHz to $2.48$GHz with $1$MHz wide bands.}}.
The attacked encryption algorithm is a software implementation of AES-$128$, whose encryption on the considered microcontroller takes $870 \mu s$.
In the following, we describe the steps used in the experiments to collect traces. 
One \emph{trace} corresponds to the collected leakage signal produced by one \ac{cp} execution.

\subsection{Leakage collection}
\begin{figure}[tb]
\centerline{\includegraphics[width=0.5\textwidth]{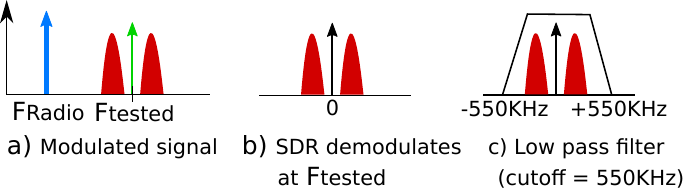}}
\caption{
\textbf{Signal processing steps for leakage collection:}
a) The SDR device collects the leakage at the targeted frequency, potentially containing leakage.
b) Trace segmentation is processed on the demodulated signal. 
c) After being segmented, the leakage is low-pass filtered at 550KHz to reduce noise.}
\label{signal_processing}
\end{figure}

Fig.~\ref{signal_processing} shows the steps undergone by the leakage signal. First, the USRP demodulates the \ac{rf} signal at a given frequency, $Ftested$, that potentially carries leakage. 
Then, the USRP samples the baseband signal at 5 MHz. 
The choice of this sampling frequency is based on previous works on screaming-channel attacks~\cite{camurati_ScreamingChannelsWhen_2018, camurati_UnderstandingScreamingChannels_2020, wang_FarFieldEM_2020},  
which we also use as it has so far provided sufficient resolution for successful attacks. 
Although a study of the impact of sampling frequency on screaming-channel attack success could be an interesting subject, it is out of the scope of this paper.

\subsection{Trace segmentation}
To segment the obtained raw traces\footnote{A \emph{raw trace} corresponds to the collected signal, sampled and quantized by the \ac{sdr}.}, i.e., to separate the segments corresponding to each individual AES encryption, pattern recognition is used. It consists in identifying the locations within the raw trace matching with the shape of the leakage produced by \acp{cp}.  
The steps applied for pattern recognition, chosen empirically for their good performance on the problem at hand, are the following:
\begin{enumerate} 
\item Low-pass filter the raw trace with a cutoff frequency at the sampling frequency divided by 4: $5\text{MHz} / 4 = 1.125\text{MHz}$.
\item Compute the sliding correlation between the pattern and the filtered trace.
\item The peaks obtained during the sliding correlation are expected to correspond to the locations of the AES segments. Segment the raw trace by cutting it at these locations.
\item To reduce noise, low-pass filter the obtained segments with a cutoff frequency of $550$KHz. 
\end{enumerate}

\begin{figure}[tb]
\centerline{\includegraphics[width=0.65\textwidth]{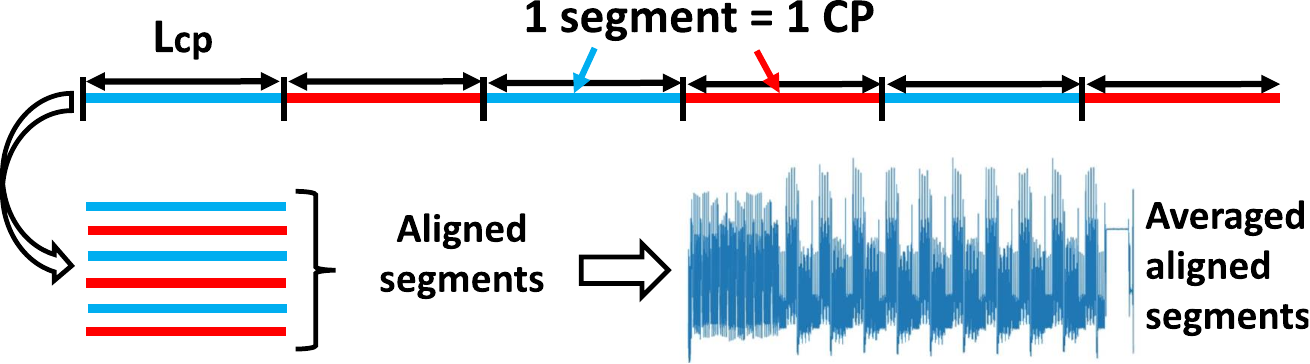}}
\caption{\textbf{\acf{vt}~\cite{guillaume_VirtualTriggeringTechnique_2022}}: Knowing the precise duration between 2 \ac{cp} executions $L_{cp}$ enables the segmentation of a raw trace. Splitting this raw trace in segments whose starts are separated by $L_{cp}$ returns segments all containing leakage of one \ac{cp} execution.
Averaging these segments returns a reduced-noise segment, which can be used as a pattern}
\label{Fig_VT}
\end{figure}

To extract an initial pattern, we use \ac{vt}~\cite{guillaume_VirtualTriggeringTechnique_2022} as the pattern extraction technique, illustrated in Fig.~\ref{Fig_VT}. 
This study shows that by knowing the precise time duration $L_{cp}$ between 2 \acp{cp}, it is possible to segment a raw trace containing leakage produced by a series of \ac{cp} executions. 
Averaging the obtained segments returns a representative pattern of the leakage produced by that device each time it executes a \ac{cp}.
The study also describes a procedure to find this precise length $L_{cp}$.

\subsection{Time diversity} 
As in previous works on screaming-channel attacks, time diversity is used to reduce noise from the leakage collection. It consists in running $N$ encryptions with exactly the same data (plaintext and key) and averaging their leakage. Since the $N$ encryptions are computing the same data, the leakage they produce would be very similar, and hence averaging tends to cancel the random noise contributions. 
While also typical in regular side-channel attacks, this is especially important in the case of screaming-channel attacks, 
as the leakage captured by the attacker contains additional noise due to the transmission channel. 

In the rest of this paper, this reduced-noise segment is called \emph{a trace}. The number $N$ is set to $10$ in the experiments where the leakage is collected through a cable (Sections~\ref{sec:Method_1_reults} and \ref{sec:Attack_non_harmonics_wired}). It is increased to $50$ for the experiments where it is collected at a distance with an antenna (Section~\ref{sec:Attack_challenging}).
These values significantly differ from the $500$ traces used in\cite{camurati_UnderstandingScreamingChannels_2020} when attacking at a distance. While this brings additional difficulty in running a successful attack, it allowed us to run the experiments in a more reasonable time of 6 days instead of 5 weeks. 
For each experiment, the number of traces collected will be noted as $Nb_{Traces} \times N_{Time\_Diversity}$.

\section{Searching for leakage at non-harmonic frequencies} 
\label{sec:SearchingLeakage}
\begin{figure}[tb]
\centerline{\includegraphics[width=0.85\textwidth]{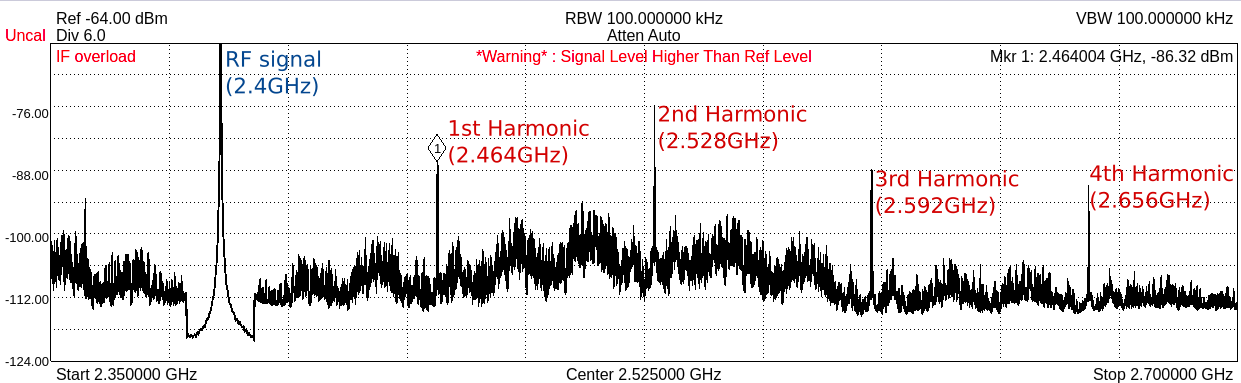}}
\caption{
\textbf{Frequency spectrum of the victim output:}
The highest peak on the left (blue text) is the legitimate \ac{rf} signal at $2.4$~GHz. The following peaks (red text) are the first four leakage harmonics at $2.464$~GHz, $2.528$~GHz, $2.592$~GHz, and $2.656$~GHz.
}
\label{Spectrum_RF_Board}
\end{figure}

Fig.~\ref{Spectrum_RF_Board} shows a part of the frequency spectrum at the output of the victim board while transmitting a Bluetooth signal at $2.4$~GHz. Next to the legitimate signal peak, other peaks with lower energy appear. These correspond to the first $4$ harmonics of the leakage from the digital part. They are present at frequencies equal to $2.4$~GHz plus multiples of $64$~MHz, which is the digital clock frequency.
Therefore, one would intuitively assume that the leakage from the digital part transmitted by the victim board is stronger at these harmonics and, hence, that it would be more challenging to perform the attack at other frequencies. 
For this reason, previous works on screaming-channel attacks~\cite{camurati_UnderstandingScreamingChannels_2020, wang_FarFieldEM_2020} use harmonic frequencies for the attack.
Typically, the second harmonic is used, as it is often less polluted by interfering signals. In contrast, the first harmonic is located within a frequency band typically used by signals like Bluetooth or WiFi.
Nevertheless, it can be seen on the spectrum that some variations in energy are also present between these harmonics, suggesting that leakage could potentially be distributed continuously along the spectrum. 

\begin{figure}[tb]
\centerline{\includegraphics[width=1\textwidth]{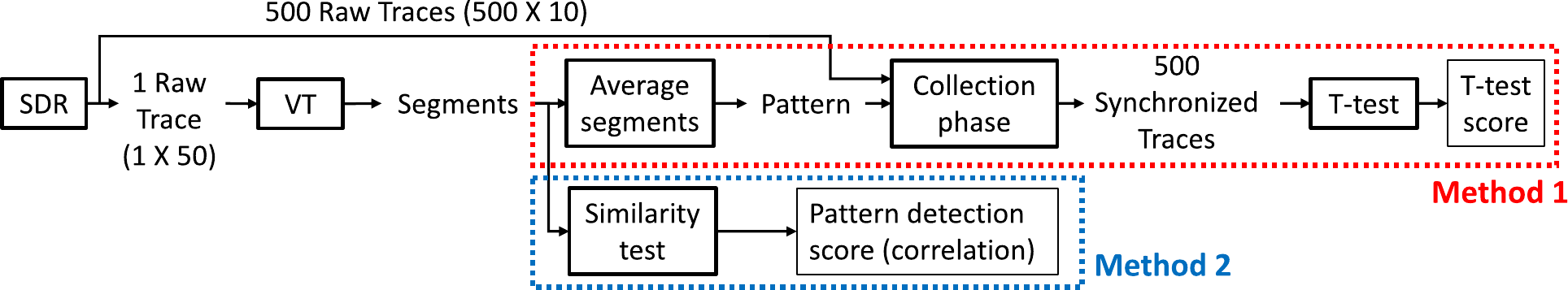}}
\caption{\textbf{Comparison of the steps of the two methods.} Both first segment a raw trace using \ac{vt}. Then, the first builds a pattern used for the collection phase and performs a t-test with the $N$ collected traces (in the presented experiment, $N=500$). The second uses the segments returned by \ac{vt} to test if the raw trace contains leakage.}
\label{approaches}
\end{figure}

The first question we want to answer is: \textbf{is the leakage from the digital part only present at the harmonic frequencies, or does it also appear at other frequencies?}
To answer this question, we propose two methodologies to investigate at which frequencies the leakage exists in the spectrum. The first involves running a t-test at each tested frequency.
The side-channel community commonly uses this test to determine whether the internal data computed by the \acp{cp} impacts the leakage.
Then, we propose a second method that reduces the implementation complexity while giving similar results.
It is an adaptation of the method used in~\cite{guillaume_VirtualTriggeringTechnique_2022} to extract the \ac{cp} leakage pattern; we refer to this second method as \emph{pattern detection}.
It consists of analyzing whether a signal collected at a given frequency is good enough to extract a \ac{cp} pattern; if yes, this means that leakage is present.
Fig.~\ref{approaches} illustrates the difference between the two methods, described in the remainder of this section.
Compared with the first methodology, the second method removes the most time-consuming phase: the collection of (500) synchronized traces needed for the t-test.

\subsection{Leakage localization using t-test method}  
\label{sec:ttest}
In this first investigation, a fixed vs. fixed t-test is performed~\cite{durvaux2016improved} at each tested frequency. 
This test indicates whether or not some leakage samples depend on the internal data computed by the \acfp{cp}. 
Therefore, if the t-test is conclusive at a given frequency, i.e., the score is over 4.5\footnote{This score means that there is information leakage with confidence $>0.99999$~\cite{gilbertgoodwill_TestingMethodSidechannel_2011,durvaux2016improved,schneider2016leakage}}, this means leakage is present there, as it is a necessary condition for the t-test to detect a dependency.
We chose to perform a fixed vs. fixed t-test as Durvaux et al.\cite{durvaux2016improved} demonstrated that it needs fewer traces to detect data dependencies compared to the classical fixed vs. random \ac{tvla} test~\cite{gilbertgoodwill_TestingMethodSidechannel_2011}.

To perform this test, two sets of \acp{cp} are executed with unique plaintext and key per set.
All \acp{cp} in one set are computed with the values of their respective set.
The leakage generated during the \ac{cp} executions is collected. 
Using leakage samples from a common time point within \acp{cp}, i.e., leakage samples generated by the same operations, we compute a t-test as indicated in the following Eq.~\eqref{eq_ttest}:

\begin{equation}
\text{t-test} = \frac{\text{û}_1 - \text{û}_2}{\sqrt{\frac{\sigma_1^{2}}{N_1} + \frac{\sigma_2^{2}}{N_2}}}
\label{eq_ttest}
\end{equation}

\noindent where $\text{û}_i$ represents the average of the samples belonging to the $i_{th}$ set, $\sigma_i^{2}$ is their variance, and $N_i$ the number of samples in this set.
Thus, for the t-test to work, it is necessary to synchronize the leakage collection with the execution of \acp{cp} perfectly 
to know which samples correspond to which time point from one leakage collection to another. The leakage collection method in section~\ref{sec:AttackScenario} is used for this. For pattern recognition, a different pattern must be extracted for each frequency, as the shape of the \ac{cp} leakage differs among frequencies. The \ac{vt}~\cite{guillaume_VirtualTriggeringTechnique_2022} technique is used to automate pattern extraction. 

In the experiment we collect $500$~traces at each frequency using the wired setup from Fig~\ref{fig:wired_setup}.
This methodology is used over the frequency range 1.4GHz to 3.4GHz ($2.4$~+/-~$1$~GHz), with a resolution of 1 MHz. These parameters are application-dependent and can be adjusted according to each particular use case to test a wider frequency band or to change the resolution. 

\subsubsection{Experimental results} 
\label{sec:Method_1_reults}
\begin{figure}[tb]
\centerline{\includegraphics[width=1\textwidth]{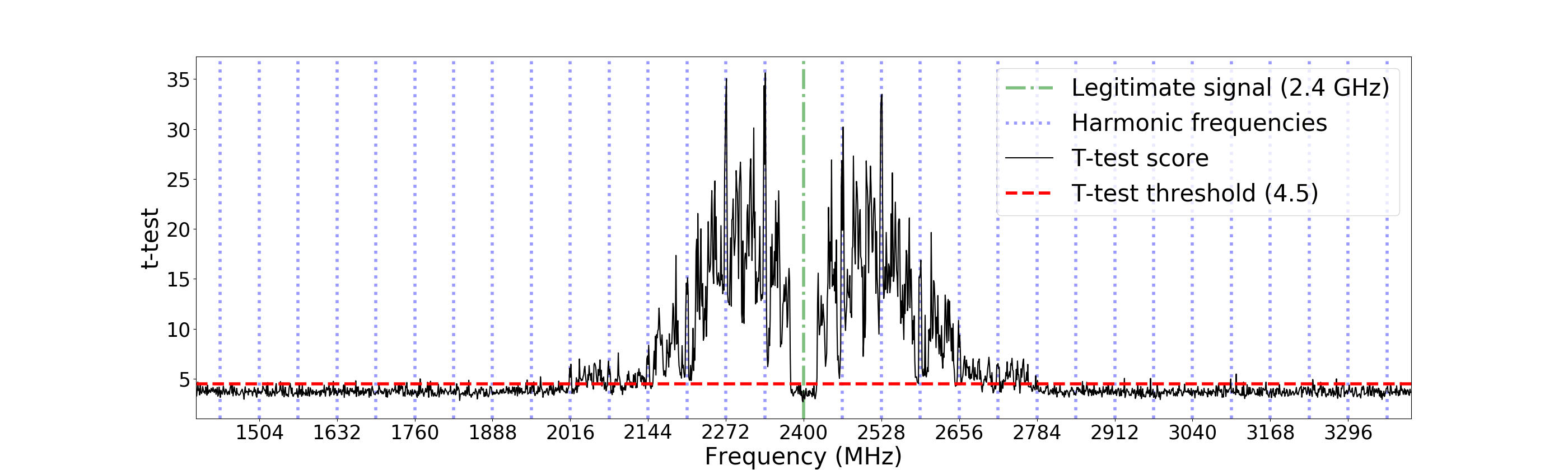}}
\caption{\textbf{Results of t-test (black curve)}: frequencies with a score higher than 4.5 (red horizontal dashed line) are considered as the ones where the leakage has been detected. Green central dash-dotted line: frequency of the legitimate signal ($2.4$~GHz). Blue vertical dotted lines: position of harmonic frequencies.}
\label{ttest}
\end{figure}

Fig.~\ref{ttest} shows the results of the t-test. At each frequency, the maximum absolute value of the t-test is kept as a score.
Frequencies corresponding to the harmonics are highlighted in blue.
If the leakage had only been present at these frequencies, the score would have been higher than this threshold only at these locations. Still, it can be observed that the score is also above 4.5 at other frequencies, suggesting that leakage is also present at non-harmonics.
In fact, even if the highest peaks are at the first $2$~harmonics, the score is very high around the first $3$~harmonics and is above the threshold over a band of more than $500$~MHz of width (almost until the $6th$~harmonic). This is true on both the right and left sides of the spectrum. 

The advantage of employing this t-test method to localize leakage in the spectrum is that it gives a result that we know how to interpret, as the t-test is a well-known tool in the side-channel community.
The downside of this method is that the collection phase takes 27 hours. A way to reduce this time would be to collect fewer traces at each frequency. 
As introduced, these results were obtained using $500$~traces at each frequency sub-band ($250$~traces per set). If we repeat the experiment for $300$ traces,  
the time is reduced to 15 hours.
However, $16.89\%$ of the frequencies previously identified as carrying leakage with the experiment using $500$~traces are not detected anymore.
Then if this method is used with an insufficient number of traces, there is a risk of not detecting frequencies that actually carry leakage.
For example, in our experiments, frequencies between the $5th$ and $6th$ harmonics, which already have a low score with $500$ traces per frequency, are not detected anymore by the t-test when it is performed with only $300$ traces. 
It is, in fact, very difficult to determine how many traces are enough because the exploitable frequencies are unknown, and hence we cannot know if some are missing. We limit the number of traces to $500$ to maintain the experiment time tractable. However, there is no guarantee that some frequencies, that could in fact contain exploitable leakage, remain undetected.

Therefore, in the next section, we propose a methodology that is more adapted to our requirements: detecting the presence of leakage in as many frequencies as fast as possible. 
The objective of this original method is to reduce the test complexity and, thus, its processing time. 
We apply this methodology to the same experiment and demonstrate equivalent results.

\subsection{Leakage localization using pattern detection method}
\begin{figure}[tb]
\centerline{\includegraphics[width=0.8\textwidth]{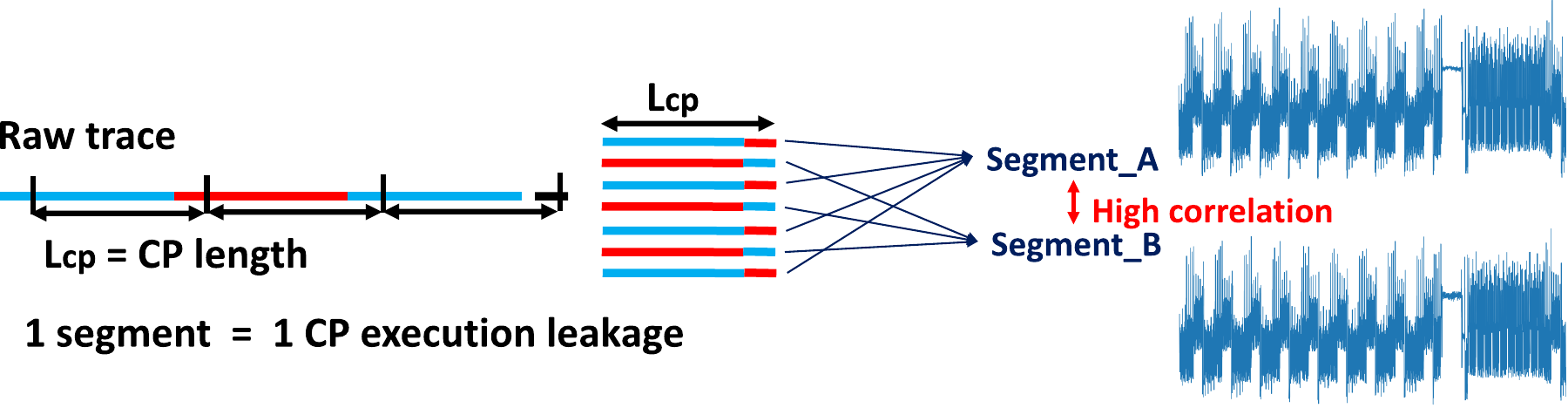}}
\caption{\textbf{Pattern detection method} 
based on \acf{vt}, which consists in segmenting a raw trace by knowing the precise time duration $L_{cp}$ between $2$ \acp{cp}. If the raw trace is collected at a frequency actually containing leakage, the obtained segments should correspond to leakage coming from the \ac{cp} and then be similar to each other, i.e., have the same shape.
}
\label{Pattern_detection}
\end{figure}

This method consists of testing the similarity of segments returned by \ac{vt} (c.f. section~\ref{sec:AttackScenario}). 
If leakage is present in the raw trace collected at the tested frequency $F$, then these segments should all correspond to the leakage of one \acf{cp} execution and have the same shape, so the similarity test should be conclusive.
We evaluate segment similarity with an adaptation of \acf{vt}~\cite{guillaume_VirtualTriggeringTechnique_2022} as illustrated in Fig.~\ref{Pattern_detection}.

We refer to this test as \emph{pattern detection}.
The algorithm of the pattern detection method is formalized in~Algorithm~\ref{alg1}. 
First, a raw trace is collected at the tested frequency $f$ (line 6). This trace is segmented using \ac{vt} (line 7), it is cut in $Nsegs$=50 segments, each segment with a size $L_{cp}$ equal to the sampling frequency$\times$the time duration between 2 \acp{cp}. The similarity test between the resulting segments is applied (lines 8 to 10). 
This test can be repeated $N_{tests}$~times (loop from lines 4 to 11) and the results averaged (line 12). In our experiments, $N_{tests}$ is initially set to $10$.
This method is applied over the same frequency range as in the first method. 

\begin{algorithm}[tb]
\caption{Pattern detection for leakage evaluation on a frequency range}
\begin{algorithmic}[1]
\small
    \Require    \text{   }                                                      \newline
    $L_{cp}$\hspace{3mm}    : The precise \ac{cp} length                        \newline
    $N_{segs}$\hspace{0mm}  : The number of \acp{cp} to cut in one Raw Trace    \newline
    $N_{tests}$\hspace{0mm} : The number of tests per frequency                  \newline 
    $F_{start}$, $F_{stop}$, $\delta_{F}$ : Start, Stop and Step Frequencies
    \Ensure    \text{   } 
    \newline
    $L_{presence}$: Estimation of the leakage presence at each tested frequency 
    \State $F \leftarrow F_{start} $
    \Repeat 
        \State $S \leftarrow \text{[ ]} $ \Comment{$S$: Similarity}
        \For{$i \gets 0$ to $N_{tests}$} 
            \State The victim device starts executing a series of \acp{cp}
            \State $RawTrace  \leftarrow \text{Signal collected by \ac{sdr} at the tested frequency $F$}$
            \State $Segs \leftarrow Segmentation($RawTrace$) \text{  using VT} $ 
            \State $Seg_A \leftarrow average(pair~Segs)$
            \State $Seg_B \leftarrow average(odd~Segs)$
            \State $S.append( \rho(Seg_A, Seg_B) )$   \Comment{Compute correlation $\rho$}
        \EndFor
        \State $L_{presence}(F)\leftarrow Mean( S )$ 
        \State $F \leftarrow F + \delta_{F} $  
    \Until {$\text{$f$} >= F_{stop}$}
    \State \textbf{return} $L_{presence}$
\end{algorithmic}
\label{alg1}
\end{algorithm}

\subsubsection{Experimental results and discussions} 
\label{sec:Method_2_reults}
\begin{figure}[tb]
\centerline{\includegraphics[width=1\textwidth]{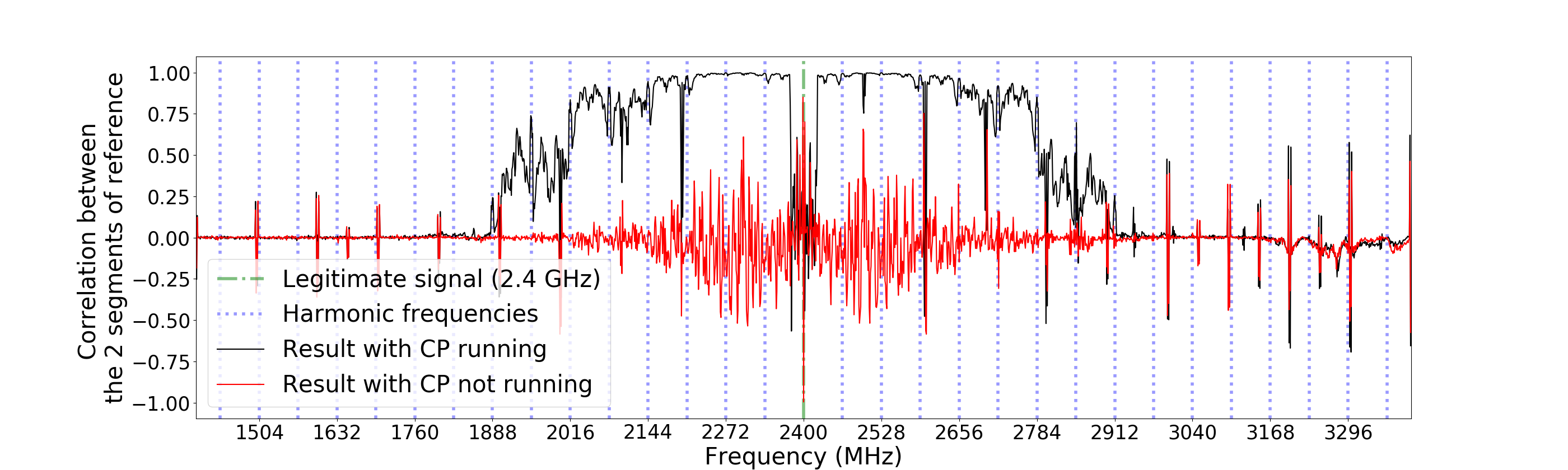}}
\caption{\textbf{Presence of leakage over the spectrum:} The graph shows the correlation between the segment of references according to the frequency where the signal was collected when \acp{cp} are running (Black curve) and not running (Red curve). 
Green Central dash-dotted line: frequency of the legitimate signal ($2.4$~GHz). Blue vertical dotted lines: position of harmonic frequencies.
}
\label{Leakage_eval}
\end{figure}

The black curve in Fig.~\ref{Leakage_eval} corresponds to the results $L_{presence}(f)$ in Algorithm~\ref{alg1}.
They are expressed as a correlation, which is representative of the similarity between segments. The higher the correlation, the more likely the \ac{cp} pattern is present.
To make sure that, as expected, the correlation is high only at certain frequencies due to the presence of \ac{cp} leakage, we repeat the same experiment, but now the victim does not perform any \acp{cp}. The red curve shows the results of this second test.
It can be seen that when \acp{cp} are not executed, there are still peaks around the carrier frequency ($2.4$~GHz) due to digital activity. However, the correlations are much weaker because when segmenting the raw traces the segments obtained are not similar as they did not correspond to \ac{cp} leakage.

\begin{figure}[tb]
\centerline{\includegraphics[width=1\textwidth]{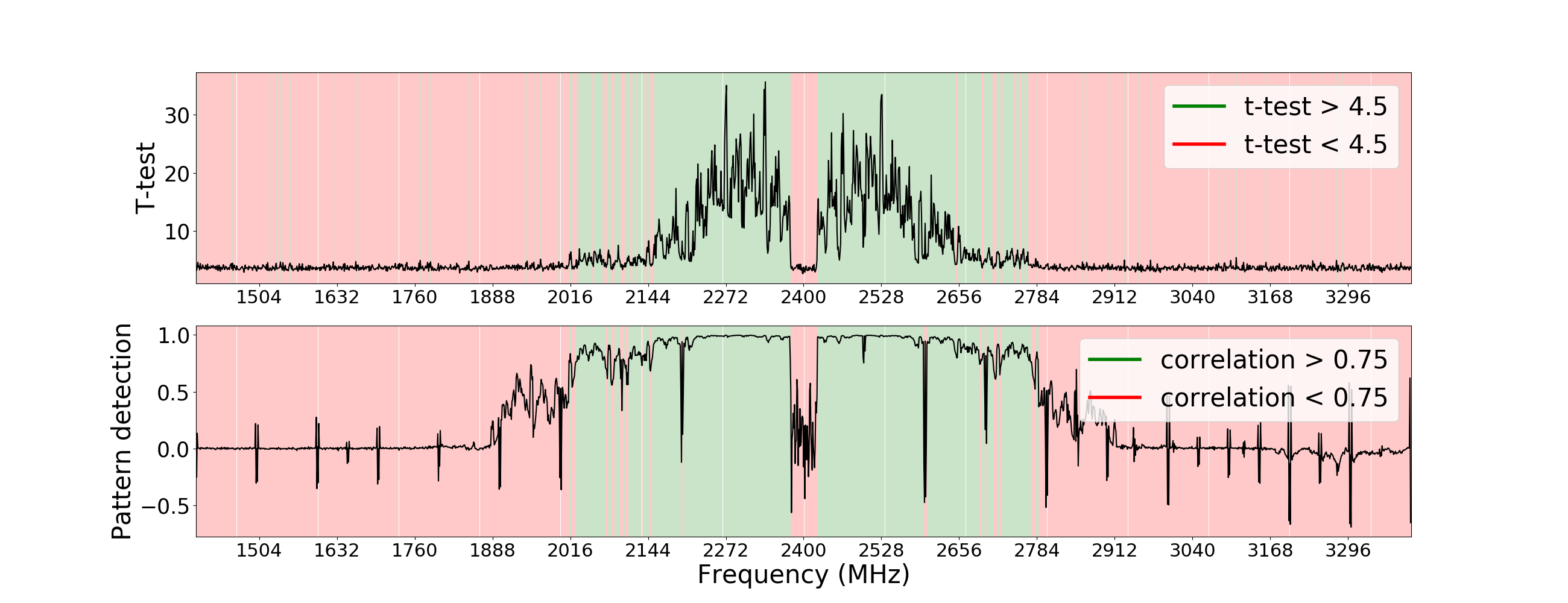}}
\caption{\textbf{Comparison of the results of the two methods:}
 The upper graph shows frequencies where \emph{t-test score $>$ $4.5$} (green underlined), hence those carrying leakage as introduced in Section~\ref{sec:ttest}. 
 The lower graph shows the results for the pattern detection method where \emph{correlation threshold $>$ $0.75$ }.
 The results obtained using the pattern detection method with this threshold are the same for $93.95\%$ of the tested frequencies.
}
\label{ttest_vs_corr}
\end{figure}

To compare both methods, we show their results side by side in Fig.~\ref{ttest_vs_corr}. 
Similarly to the threshold of $4.5$ used in the t-test method, we need to set another condition for this second method to consider the detection as positive. The comparison of both methods for a threshold of minimum correlation larger or equal to $0.75$ is shown in Fig.~\ref{ttest_vs_corr}.
We chose this threshold as it corresponds to the highest score of the test in the absence of leakage (red curve). Therefore, any score below this threshold cannot be taken as an indication of the presence of leakage.
For both methods, the frequencies where their respective condition is met are highlighted in green.
For this selected threshold of $0.75$, we found after analyzing the results for each frequency that both methods yield the same results for $93.95\%$ of the tested frequencies.

For this second method, it took 50 minutes instead of the 27 hours in method 1 of Section~\ref{sec:ttest}.
It is possible to further reduce this time by reducing the number of similarity tests performed at each frequency. As introduced, $N_{tests}$ was initially set to $10$ tests per frequency. 
We repeated the experiment reducing $N_{tests}$ to only $1$. In this case, leakage localization took only 15 minutes without significantly altering the results, and $94.78\%$ of the detected frequencies with $N_{tests}$=$10$ were still detected with $N_{tests}$=$1$.

\section{Attacking  at non-harmonics}
\label{sec:Attack_non_harmonics_wired}
In Section~\ref{sec:SearchingLeakage}, we demonstrated that the leakage is also present at non-harmonics frequencies. This brings another question: \textbf{how efficient are attacks at these non-harmonic frequencies?}
To answer the question, the attack is performed on a part of the spectrum where the leakage localization methods gave the best results.
To keep the experimentation phase tractable, experiments only cover the right-hand side of the spectrum (i.e., positive) with respect to the legitimate signal. Indeed, the objective of the experiment is not to evaluate all possible frequencies on this particular type of device but to determine if an attack is possible at non-harmonic frequencies and to compare their performance to attacks at harmonic frequencies. The experiment covers a range of frequencies from $2.45$~GHz to $2.6$~GHz, and attacks are centered at $150$ different frequencies, with 1MHz steps.
One may note that the attack would probably also be possible at other frequencies, including the one on the left part of the spectrum. However, as indicated, we focus only on the right half side to keep the experimentation phase in a reasonable time.

\subsection{The attack and score}
In our experiments, we run profiled correlation attacks\cite{durvaux2016improved}, where the attacker has access to a similar device as the victim. This enables the attacker to build a profile for this type of device and learn the leakage behavior. 
During the attack phase, for each key byte, an assumption (i.e., hypothesis) is made on their value.
The 256 possible hypotheses are tested, each getting a probability to be the correct one based on the correlation between the estimated leakage using the profile on one side, and the real leakage produced by the victim on the other. 
The hypothesis giving the highest probability is assumed to be the correct one.

In many cases, some bytes are incorrectly guessed, 
but it is still possible to brute-force the correct key. 
A brute-force approach~\cite{poussier2016simple} consists in testing the ranked keys from the most probable, according to the probabilities computed during the attack, until finding the right one. 
The number of keys tested by the brute-force algorithm before reaching the correct one is the \emph{Key Rank} and is representative of the complexity of recovering the key. 
The lower the key rank is, the better the attack performs, as the brute-force attack needs less time to reach the correct key.
When the key rank is lower than $2^{32}$, it takes about $5$~minutes on the experimental computer to brute-force the key. When lower than $2^{35}$, the brute-force takes about $1$~hour.
In the remainder of this paper, this key rank is kept as the criteria to evaluate the efficiency of an attack as in previous works \cite{camurati_UnderstandingScreamingChannels_2020}.

\subsection{Experiment and results}
\label{sec:ExpCable}
In this first experiment, the victim and attacker are still connected by a cable 
(Fig.~\ref{fig:wired_setup}).
Two sets of $15000\times10$ traces are collected at each frequency, one to build the profile and the other to test the attack. The collection phase takes $4$ days.
For each attack, we compute the key rank and show the results in Fig.~\ref{fig:Fig_BCs_15000T_cable}. 
The experiment confirms that the attack is not only succeeding at the harmonics as expected but also at many other frequencies.
Then, this finding increases the number of potential frequencies to use to succeed in the attack. The key rank is lower than $2^{32}$ at the $3$ harmonics as they are not polluted. Among the 147 non-harmonics, 105 have a key rank lower than $2^{32}$ and 12 lower than $2^{35}$.
       
As the experiment is fully automated, it is important to notice that a very high key rank at a given frequency does not ensure the attack is necessarily more difficult there. 
But it can be tried to make it work better by putting more effort into it, which is not our concern here.

\section{Attacking in challenging conditions}
\label{sec:Attack_challenging}
After demonstrating attacks are also possible at frequencies other than the harmonics, we investigate how useful this finding is in a noisy environment when attacking at a distance.
The questions targeted by this second experiment are the following: 
\textbf{In a noisy environment where harmonics are polluted, can non-harmonic frequencies keep the attack feasible?}
\textbf{Can the attack be better at non-harmonic frequencies than at harmonic frequencies}

In the experiments presented in this section, the leakage is collected using the antenna and the setup shown in Fig.~\ref{fig:setup_7meters}.
Compared to the first experiment we increase time diversity from $10$ to $50$, but collect the same number of traces at each frequency ($15000\times50$).
The patterns used for the collection phase and profiles used for the attack are the ones that were built in~Section~\ref{sec:ExpCable}.
The key rank is kept as the attack score.

\subsection{Attacking at a distance in a noisy environment}
A first test is performed with the antenna at 2 meters.  
In these conditions, the collection phase takes six days. 
Fig.~\ref{fig:Fig_BCs_15000T_2meters} shows the results. 
As expected, we can observe how the noisy environment reduces the number of exploitable frequencies. This is particularly visible around the first harmonic at $2.464$~GHz, where WiFi and Bluetooth signals are present.
Among the $150$ frequencies, the rank is lower than $2^{32}$ only at $2$ harmonics, as the first one is polluted. Among the $147$ non-harmonics, $78$ have a rank lower than $2^{32}$ and $12$ lower than $2^{35}$.

\subsection{Attacking with fewer traces}
A common goal of side-channel attacks is to succeed with as few traces as possible.
We re-computed the attack with the traces collected at 2 meters but reduced the number of traces per attack to $750\times50$ (from $15000\times50$). We were then able to set up ($15000 / 750 = 20$~attacks at each frequency). The experiment provides the results shown in Fig.~\ref{fig:Fig_BCs_750T_2meters}
After sorting the $150$ frequencies according to the average of their scores (average of the $20$ log2(key rank)), the harmonics ranked at the $3$\textsuperscript{rd}, $7$\textsuperscript{th} and $121$\textsuperscript{st} place. This proves that some non-harmonics can even get better scores than harmonic frequencies.

\subsection{Attacking at a further distance}
\begin{figure}[!tb]
    \centering
    \begin{subfigure}{1\textwidth}
        \centering
        \includegraphics[width=\textwidth]{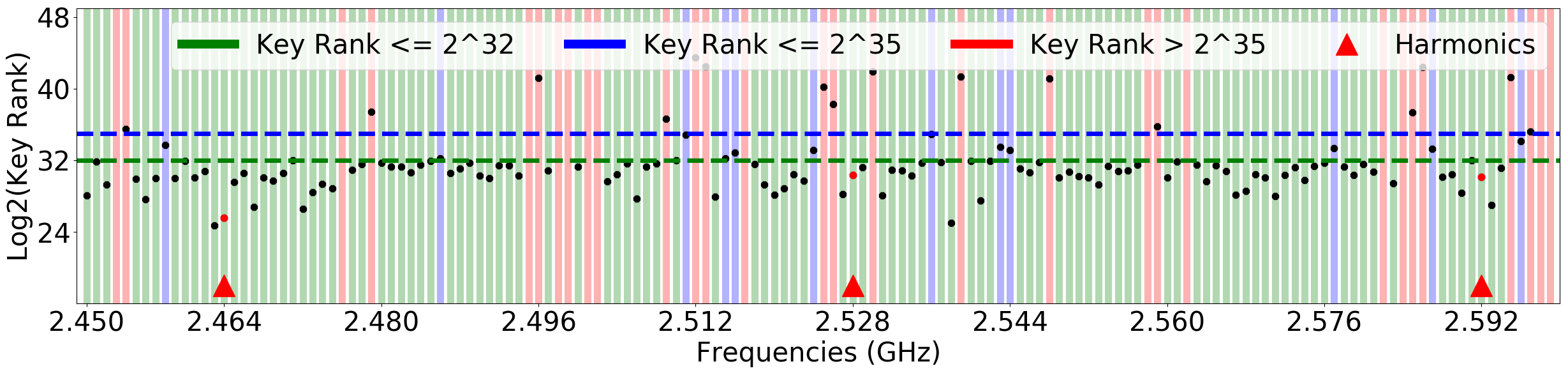}
        \caption{Attack by cable using $15000$ traces}
        \label{fig:Fig_BCs_15000T_cable}
    \end{subfigure}
    \hspace{6ex}
    \begin{subfigure}{1\textwidth}
        \centering
        \includegraphics[width=\textwidth]{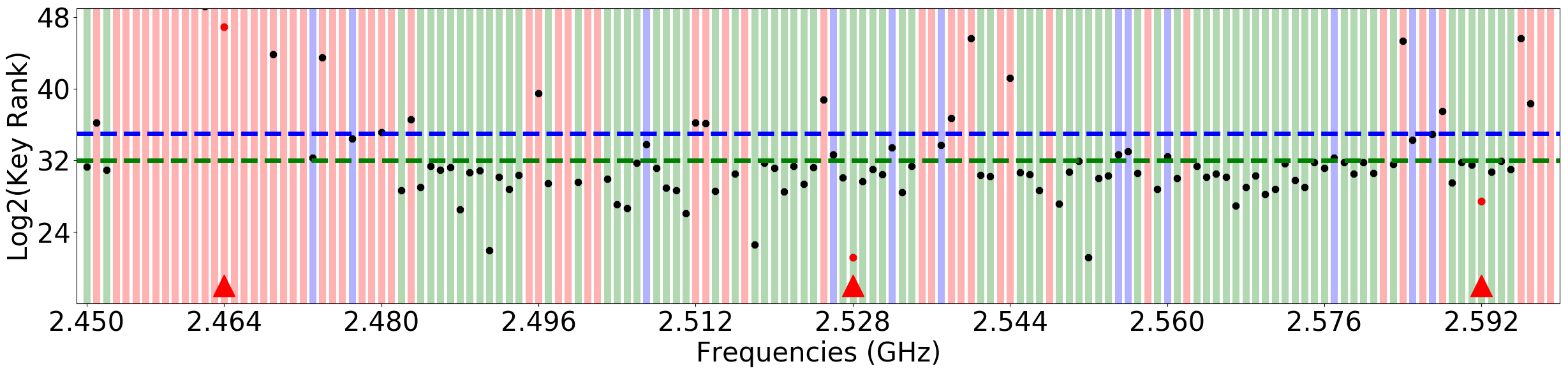}
        \caption{Attack at 2 meters using $15000$ traces}
        \label{fig:Fig_BCs_15000T_2meters}
    \end{subfigure}
    \hspace{6ex}
    \begin{subfigure}{1\textwidth}
        \centering
        \includegraphics[width=\textwidth]{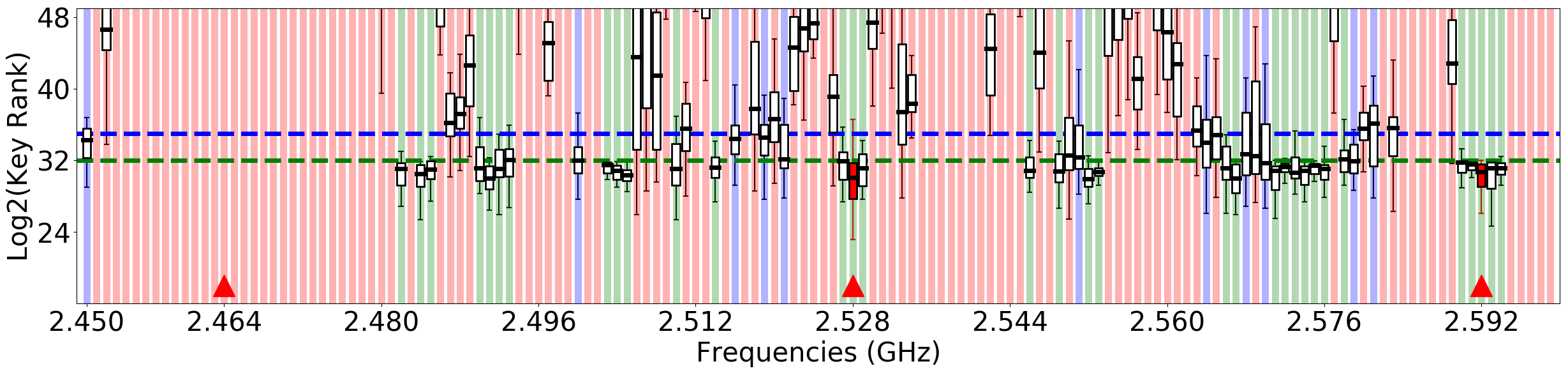}
        \caption{Attack at 2 meters using 750 traces (20 attacks per frequency)}
        \label{fig:Fig_BCs_750T_2meters}
    \end{subfigure}
    \hspace{6ex}
    \begin{subfigure}{1\textwidth}
        \centering
        \includegraphics[width=\textwidth]{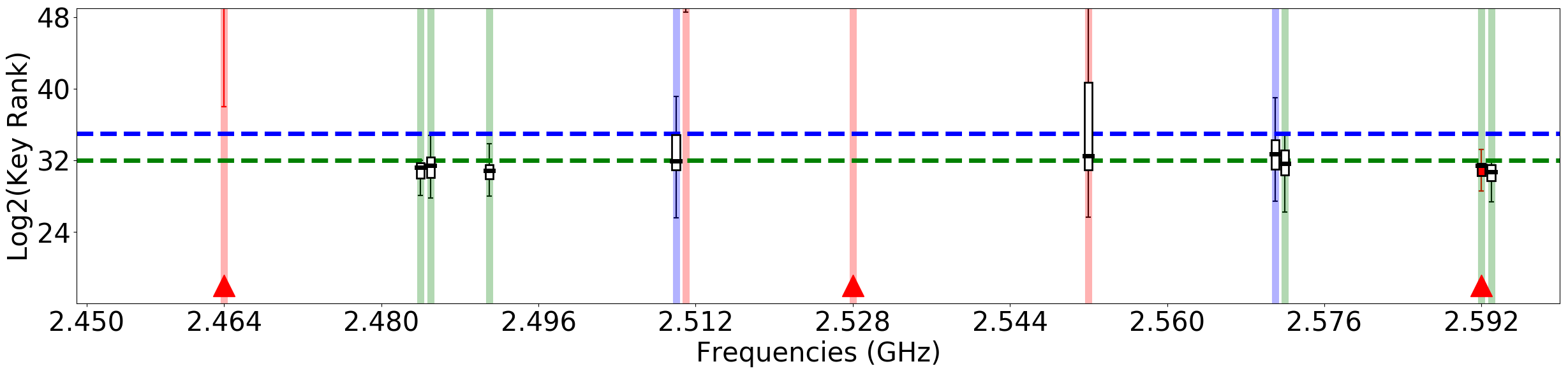}
        \caption{Attack at 7 meters using 750 traces (50 attacks per frequency)}
        \label{fig:Fig_BCs_750T_7meters}
    \end{subfigure}
    \caption{\textbf{Experimental results:} Key Rank (the lower, the better). Underlying colors per frequency indicate key ranks of: green $<=$ to $2^{32}$ (approx. $5$~mins brute-force), blue $<=$ to $2^{35}$ (approx. $1$~hour brute-fore), red $>$ to $2^{35}$. 
For (\subref{fig:Fig_BCs_750T_2meters}) and (\subref{fig:Fig_BCs_750T_7meters}), the graph shows: (1) a central box with the distribution of scores from the first quartile to the third; (2) whiskers with the extension of the remaining scores by $1.5\times$ the inter-quartile range (equal to the difference between the third and first quartile); and (3) a horizontal line with the median score of the attacks. 
} 
\label{Fig_BCs_challenging}
\end{figure}

As screaming-channel attacks try to enable attacks where attackers are as far as possible from the victim, we run a new round of attack experiments with the antenna put at a distance of $7$ meters.
This time $50$\footnote{$50$ is the minimal number usually considered by the side-channel community for statistically meaningful results} attacks are performed under the exact same conditions at each frequency.
To keep the experiments feasible in a reasonable time, we reduced the number of tested frequencies.
The frequencies selected for this attack at $7$ meters are chosen among the ones where the attack performed best in the 2-meters scenario using $750$~traces. 
The results of the attacks at $7$~meters are shown in Fig.~\ref{fig:Fig_BCs_750T_7meters}. 
In this case, there is \textbf{only one harmonic that still gets a key rank lower than $2^{35}$, while most of the selected non-harmonics still work}. Among the $9$ which were selected, $5$ have a lower rank than $2^{32}$ and $2$ than $2^{35}$.

\begin{figure}[!tb]
    \centering
    \begin{subfigure}{0.495\textwidth}
        \centering
        \includegraphics[width=\textwidth]{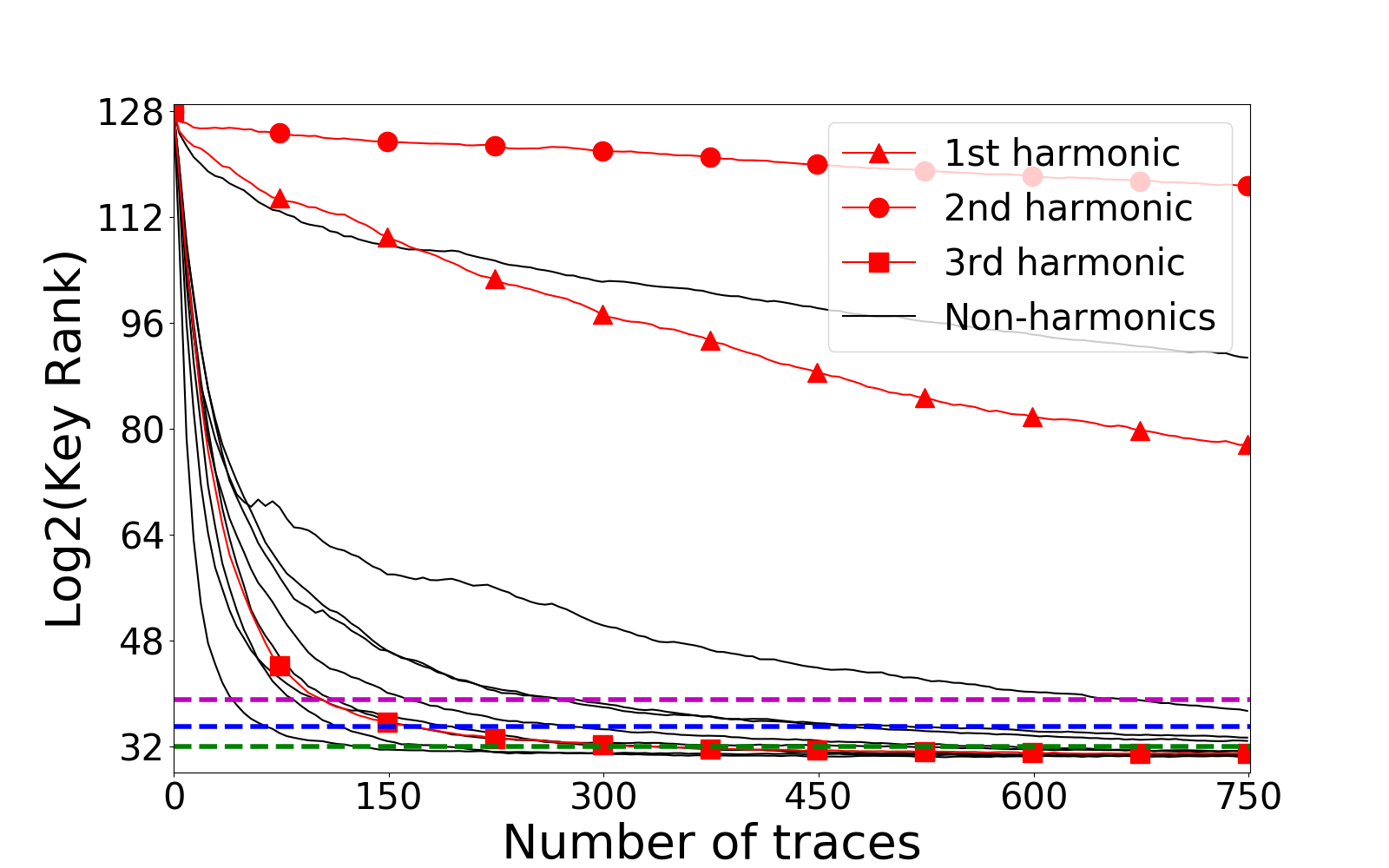}
        \caption{}
        \label{fig:BCs_NbTraces_a}
    \end{subfigure}
    \begin{subfigure}{0.495\textwidth}
        \centering
        \includegraphics[width=\textwidth]{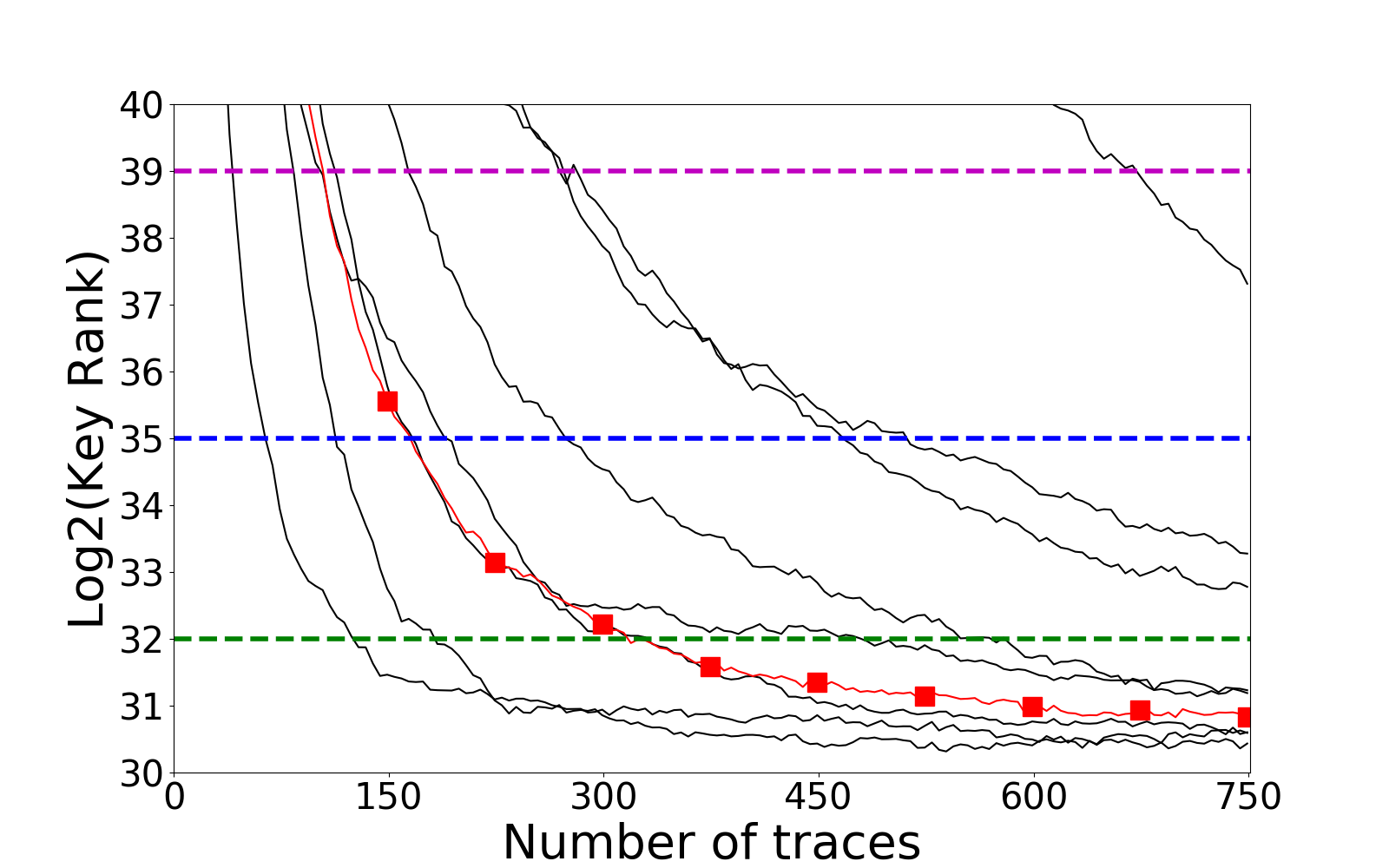}
        \caption{}
        \label{fig:BCs_NbTraces_b}
    \end{subfigure}
    \caption{\textbf{Attack at 7 meters:} a) Key Rank (the lower, the better) according to the number of traces used for the attack. b) Same results with reduced scale.}
    \label{BCs_NbTraces}
\end{figure}

Fig.~\ref{BCs_NbTraces} shows the results of the same attacks, but it focuses on the evolution of the key rank according to the number of traces used. 
Again, we keep the average of the $50$ log2 (key rank) as the attack score for each frequency.
When using the same harmonic as in previous works~\cite{camurati_ScreamingChannelsWhen_2018, camurati_UnderstandingScreamingChannels_2020, wang_FarFieldEM_2020} (the second one at $2.528$~GHz), the key rank decreases but very slowly.
Then in our case, to get a rank lower than $2^{35}$ using this frequency, it is necessary to collect up to $30286\times50$ traces. 
These results show how allowing to search for leakage at frequencies other than the harmonics considerably reduces the number of traces needed to get the same results.
The best non-harmonic frequency, at $2.484$~GHz, needs only $65\times50$ traces to get this result, which is even better than the best harmonic at $2.592$~GHz, where the number of traces required is $166\times50$.

\section{Discussion and conclusion}
\label{sec:Conclusion}

This work defied the assumption that screaming-channel attacks perform best (or only) at harmonics of the digital processing clock of the victim, frequencies where the leakage was so far supposed to be present with the highest amplitude.
To investigate this, we proposed two methods to locate and evaluate leakage over a band of the frequency spectrum.
The first method, the most intuitive and direct, builds from the literature on side-channel attacks and tries to find exploitable leakage through a t-test at each tested frequency. We used a fixed vs. fixed test due to its better performance with fewer traces.
The second method is an original contribution of this work that tries to reduce the implementation complexity and the processing time while keeping the same quality of results as the first method, which uses a standard methodology accepted by the side-channel community.
Exploiting these two methods, we demonstrate that the leakage is also present at a large amount of non-harmonic frequencies.

The presence of leakage at non-harmonics is consistent with previous studies~\cite{noulis2017cmos}, demonstrating that when the digital part creates noise at a given frequency, and as this noise travels through the CMOS substrate, the latter acts as a filter that spreads the noise over a wider frequency band. As a consequence, the noise can be found on the \ac{rf} side at frequencies other than the harmonics.

We considered only one type of device, the same used by previous works on screaming-channel attacks, that is still available off-the-shelf at the time of our study.
The present study does not prove that leakage will always be present at non-harmonics on any other device. However, it highlights the fact that leakage presence has to be checked there, too, as it is possible to find it at these frequencies, even if stronger peaks at harmonics give the intuitive idea that leakage would appear mainly there. This is exactly what we have proved in this work.

This study also demonstrates how this phenomenon can make attacks feasible in cases where all exploitable harmonics are polluted by interfering signals, as is the case in more realistic, real-life scenarios. 
The studied phenomenon can also reduce the number of traces needed for the attack. 
Compared with the performance of the attack at the best harmonic, using the best non-harmonic enables to reduce by $60\%$ the number of traces needed to get a key rank under $2^{35}$.

In future works, it could be interesting to detect the best frequencies first and then focus the efforts only on them. For example, by building better profiles: in our work, in order to build profiles at a large number ($150$) of frequencies, these profiles were built with only $150$K traces ($15000\times10$), which is relatively small compared with previous works ($1$ to $5$M traces per profile).
One can also extend the range of attacks to a distance where no harmonic gives a reasonable key rank (for example, superior to $2^{39}$) with a given maximum number of traces and observe how many meters a non-harmonic attack can gain.

\section*{Acknowledgment}
We want to acknowledge the reviewers of the current and previous versions of this paper, as well as Dr. Maria Méndez Real and Dr. Dennis Gnad for their constructive feedback.

\bibliographystyle{splncs04}
\bibliography{biblio}

\begin{thebibliography}{10}
\providecommand{\url}[1]{\texttt{#1}}
\providecommand{\urlprefix}{URL }
\providecommand{\doi}[1]{https://doi.org/#1}

\bibitem{adamczyk2017foundations}
Adamczyk, B.: Foundations of electromagnetic compatibility: with practical applications. John Wiley \& Sons (2017)

\bibitem{afzali2006substrate}
Afzali-Kusha, A., Nagata, M., Verghese, N.K., Allstot, D.J.: Substrate noise coupling in soc design: Modeling, avoidance, and validation. Proceedings of the IEEE  \textbf{94}(12),  2109--2138 (2006)

\bibitem{agrawal_EMSideChannel_2003}
Agrawal, D., Archambeault, B., Rao, J.R., Rohatgi, P.: The {{EM Side}}\textemdash{{Channel}}(s). In: Cryptographic {{Hardware}} and {{Embedded Systems}} - {{CHES}} 2002. pp. 29--45. {Springer} (2003)

\bibitem{Agrawal_multichannelAttack_2003}
Agrawal, D., Rao, J.R., Rohatgi, P.: Multi-channel attacks. In: Cryptographic Hardware and Embedded Systems--CHES 2003. pp. 2--16. Springer (2003)

\bibitem{brier_CorrelationPowerAnalysis_2004}
Brier, E., Clavier, C., Olivier, F.: Correlation power analysis with a leakage model. In: Cryptographic Hardware and Embedded Systems--CHES 2004. pp. 16--29. Springer (2004)

\bibitem{camurati_UnderstandingScreamingChannels_2020}
Camurati, G., Francillon, A., Standaert, F.X.: Understanding screaming channels: From a detailed analysis to improved attacks. IACR transactions on cryptographic hardware and embedded systems pp. 358--401 (2020)

\bibitem{camurati_ScreamingChannelsWhen_2018}
Camurati, G., Poeplau, S., Muench, M., Hayes, T., Francillon, A.: Screaming channels: When electromagnetic side channels meet radio transceivers. In: ACM Conference on Computer and Communications Security. pp. 163--177 (2018)

\bibitem{chari_TemplateAttacks_2003}
Chari, S., Rao, J.R., Rohatgi, P.: Template {{Attacks}}. In: Cryptographic {{Hardware}} and {{Embedded Systems}}--{{CHES}} 2002 (2003)

\bibitem{choiTEMPESTComebackRealistic2020}
Choi, J., Yang, H.Y., Cho, D.H.: {{TEMPEST Comeback}}: {{A Realistic Audio Eavesdropping Threat}} on {{Mixed-signal SoCs}}. In: Proceedings of the 2020 {{ACM SIGSAC Conference}} on {{Computer}} and {{Communications Security}} (2020)

\bibitem{dessouky_SoKSecureFPGA_2021}
Dessouky, G., Sadeghi, A.R., Zeitouni, S.: {{SoK}}: {{Secure FPGA Multi-Tenancy}} in the {{Cloud}}: {{Challenges}} and {{Opportunities}}. In: IEEE EuroS\&P. pp. 487--506 (2021)

\bibitem{durvaux2016improved}
Durvaux, F., Standaert, F.X.: From improved leakage detection to the detection of points of interests in leakage traces. In: Advances in Cryptology--EUROCRYPT 2016. pp. 240--262. Springer (2016)

\bibitem{gandolfi_EM_2001}
Gandolfi, K., Mourtel, C., Olivier, F.: Electromagnetic analysis: Concrete results. In: Cryptographic Hardware and Embedded Systems--CHES 2001. pp. 251--261. Springer (2001)

\bibitem{gierlichsMutualInformationAnalysis2008a}
Gierlichs, B., Batina, L., Tuyls, P., Preneel, B.: Mutual information analysis: A generic side-channel distinguisher. In: Cryptographic Hardware and Embedded Systems--CHES 2008. pp. 426--442. Springer (2008)

\bibitem{gilbertgoodwill_TestingMethodSidechannel_2011}
Gilbert~Goodwill, B.J., Jaffe, J., Rohatgi, P.: A testing methodology for side-channel resistance validation. In: {{NIST}} Non-Invasive Attack Testing Workshop. vol.~7, pp. 115--136 (2011)

\bibitem{guillaume_VirtualTriggeringTechnique_2022}
Guillaume, J., Pelcat, M., Nafkha, A., Salvador, R.: Virtual triggering: a technique to segment cryptographic processes in side-channel traces. In: 2022 IEEE Workshop on Signal Processing Systems (SiPS). pp.~1--6. IEEE (2022)

\bibitem{kocher_DifferentialPowerAnalysis_}
Kocher, P., Ja, J.: Differential {{Power Analysis}}. Advances in Cryptology--CRYPTO’99  (1999)

\bibitem{le2011experimental}
Le, J., Hanken, C., Held, M., Hagedorn, M.S., Mayaram, K., Fiez, T.S.: Experimental characterization and analysis of an asynchronous approach for reduction of substrate noise in digital circuitry. IEEE transactions on very large scale integration (VLSI) systems  \textbf{20}(2),  344--356 (2011)

\bibitem{mangard2008power}
Mangard, S., Oswald, E., Popp, T.: Power analysis attacks: Revealing the secrets of smart cards, vol.~31. Springer Science \& Business Media (2008)

\bibitem{masure_ComprehensiveStudyDeep_2019}
Masure, L., Dumas, C., Prouff, E.: A {{Comprehensive Study}} of {{Deep Learning}} for {{Side-Channel Analysis}}. IACR Transactions on Cryptographic Hardware and Embedded Systems  (2019)

\bibitem{mohamed2010physical}
Mohamed, C., Barelaud, B., Ngoya, E.: Physical analysis of substrate noise coupling in mixed circuits in soc technology. In: The 5th European Microwave Integrated Circuits Conference. pp. 274--277. IEEE (2010)

\bibitem{noulis2017cmos}
Noulis, T., Baumgartner, P.: {CMOS substrate coupling modeling and analysis flow for submicron SoC design}. Analog Integrated Circuits and Signal Processing  \textbf{90},  477--485 (2017)

\bibitem{poussier2016simple}
Poussier, R., Standaert, F.X., Grosso, V.: Simple key enumeration (and rank estimation) using histograms: An integrated approach. In: Cryptographic Hardware and Embedded Systems--CHES 2016. pp. 61--81. Springer (2016)

\bibitem{rhee2008experimental}
Rhee, W., Jenkins, K.A., Liobe, J., Ainspan, H.: Experimental analysis of substrate noise effect on pll performance. IEEE Transactions on Circuits and Systems II: Express Briefs  \textbf{55}(7),  638--642 (2008)

\bibitem{schellenberg_JobRemotePower_2018a}
Schellenberg, F., Gnad, D.R.E., Moradi, A., Tahoori, M.B.: An {{Inside Job}}: {{Remote Power Analysis Attacks}} on {{FPGAs}}. 2018 Design, Automation \& Test in Europe Conference \& Exhibition (DATE) p.~6 (2018)

\bibitem{schneider2016leakage}
Schneider, T., Moradi, A.: Leakage assessment methodology: Extended version. Journal of Cryptographic Engineering  \textbf{6},  85--99 (2016)

\bibitem{standaert_IntroductionSideChannelAttacks_2010}
Standaert, F.X.: Introduction to side-channel attacks. Secure integrated circuits and systems pp. 27--42 (2010)

\bibitem{wang_FarFieldEM_2020}
Wang, R., Wang, H., Dubrova, E.: Far field em side-channel attack on aes using deep learning. In: 4th ACM Workshop on Attacks and Solutions in Hardware Security. pp. 35--44 (2020)

\end{thebibliography}

\end{document}